\pdfoutput=1

\documentclass[aps,prfluids,superscriptaddress]{revtex4-2}

\usepackage[T1]{fontenc} 
\usepackage{amsmath}
\usepackage{amsfonts}
\usepackage{amssymb}
\usepackage{graphicx}
\usepackage{xcolor}

\usepackage{enumitem}

\usepackage{booktabs} 
\let\oldhat\hat

\renewcommand{\hat}[1]{\oldhat{\mathbf{#1}}}

\usepackage{multirow} 

\begin{document}


\title{Modelling the Dynamics of Partially Wetting Droplets on Fibres}



\author{Raymond Christianto}
\affiliation{Department of Physics, Durham University, DH1 3LE, United Kingdom}

\author{Yudi Rahmawan}
\affiliation{Department of Mechanical Engineering, Universitas Pertamina, Jakarta 12220, Indonesia}

\author{Ciro Semprebon}
\email[]{ciro.semprebon@northumbria.ac.uk}
\affiliation{Department of Mathematics, Physics and Electrical Engineering, Northumbria University, Newcastle-upon-Tyne NE1 8ST, United Kingdom}

\author{Halim Kusumaatmaja}
\email[]{halim.kusumaatmaja@durham.ac.uk}
\affiliation{Department of Physics, Durham University, DH1 3LE, United Kingdom}

\date{\today}

\begin{abstract}
We simulate gravity driven dynamics of partially wetting droplets moving along a fibre using the lattice Boltzmann method. For the so-called clamshell morphology, we find three possible dynamic regimes upon varying the droplet Bond number and the fibre radius: compact droplet, droplet breakup and droplet oscillation. For small Bond numbers, in the compact droplet regime, the droplet reaches a steady state and its velocity scales linearly with the driving body force. A similar scaling law can also be observed for the barrel morphology. For higher Bond numbers, in the droplet breakup regime, satellite droplets are formed trailing the initial moving droplet. We find such droplet satellite formation is easier with increasing fibre curvature (smaller fibre radius). Finally, in the droplet oscillation regime, favoured in the mid-range of fibre radius, the droplet shape periodically extends and contracts along the fibre.
\end{abstract}


\maketitle

\section{Introduction}

Fibres and fibrous materials are ubiquitous in nature and industry, and their interactions with liquid droplets are often key for their use and functions. In nature, numerous plants and animals, such as cactus and spider silk, have unique fibre-like features to condensate and collect water droplets efficiently \cite{ju2012multi,bai2012functional}. Mammalian hair fibre has a series of axially symmetrical serrations on its cylindrical main structure, directing water droplets to move parallel to the fibre \cite{carroll1989droplet}. Understanding droplet wetting on fibres is also important for small scale (laboratory) and large scale (industrial) applications. For example, one way to prepare a compound droplet -- microdroplets encapsulated by a larger droplet -- is by using fibre junctions \cite{weyer2015compound}. Fibre networks have been suggested as a possible open droplet microfluidic design \cite{gilet2009digital}. Mimicking nature, fibre mesh structures have been harnessed for improved designs of water collecting and fog harvesting devices \cite{von2015bioinspired, park2013optimal,ghosh2020influence}.

Given their wide-ranging interest, droplet studies on fibres have a long history. In a seminal paper, Lord Rayleigh has observed that a thin film encapsulating a fibre can break up because of capillary instability \cite{rayleigh1878on}, resulting in a string of droplets with barrel morphology, where the droplet is axisymmetric and encapsulates the fibre completely. Adams noted another morphology, clamshell, which is favoured by the droplet when the contact angle of the droplet increases \cite{adam1937detergent}. Clamshell is characterised by its asymmetric shape, where the droplet is only perched on one side of the fibre. More recent studies have systematically investigated the transition and the stability of these two morphologies \cite{eral2011drops,mchale2002global, chou2011equilibrium, de2012buoyant}, which depend on the contact angle of the droplet, the droplet volume, and the fibre radius.

Beyond equilibrium properties, droplet dynamics on fibres has also generated significant interest. Several groups have deduced the scaling laws for droplet velocity along the fibre by balancing the viscous forces experienced by the droplet against the driving force \cite{lorenceau2004drops, gilet2010droplets}. Typically, the droplet weight due to gravity is the driving force in the experiments. Other external driving forces, both parallel and perpendicular to the fibre, have also been considered to generate droplet motion, including subjecting the droplet to airflow \cite{mullins2006observation,bintein2019self} or using a centrifugal force by rotating the fibre support \cite{texier2015droplets}. In addition, more complex fibre geometries have been investigated. These include conical fibres \cite{ju2013bioinspired, liang2015drops, mccarthy2019dynamics, van2021capillary}, where the curvature asymmetry can be exploited to drive self-propulsion; fibre cross-junctions \cite{davoudi2016barrel, weyer2017switching}, where the droplet trajectory can be tuned, for example, as the droplet size and orientation of the driving force is varied; and multiple parallel fibres, both when the fibres are rigid and flexible \cite{sauret2015wetting, aziz2019competing, wang2021hysteresis}.

To date, the majority of works on droplet dynamics have focused on the case where the droplet is perfectly wetting the solid fibres. In contrast, here we will primarily consider the dynamics of partially wetting droplets. We will investigate the complex shape and velocity of such droplets as we vary the fibre curvature and the droplet body force (correspondingly, velocity). When the droplet remains compact, we identify scaling laws for both the barrel and clamshell morphologies. Compared to known results for fully wetting droplets \cite{lorenceau2004drops, gilet2010droplets}, a key difference is in how the loss of symmetry and corresponding variation of the droplet shape and aspect ratio affect the dynamics of droplets with clamshell morphology. Focussing on the clamshell morphology, we then demonstrate that the droplet dynamics is affected by the fibre curvature. On flat surfaces it has been observed that, with increasing velocity, the receding contact line becomes unstable leading to the formation of trailing satellite droplets. This phenomenon, the so-called pearling instability, has been extensively studied \cite{podgorski2001corners, le2005shape, peters2009coexistence, engelnkemper2016morphological}.
Interestingly, we find satellite droplet formation occurs on fibres as well, and it is favoured with increasing fibre curvature. Moreover, we find a region of the parameter space in which the droplet oscillates as it moves along the fibre.
A similar observation was made by Yang et al. \cite{yang2020effects} for droplets on an inclined flat surface.

This paper is organised as follows. In section II, we illustrate our simulation approach based on the lattice Boltzmann method \cite{kruger2016lattice}. The lattice Boltzmann method has been successfully harnessed to study a wide range of interfacial flows, including droplet collisions \cite{inamuro2004lattice, wohrwag2018ternary}, capillary filling, self-propelled droplet slugs \cite{bala2019wetting}, and droplet impact on textured superhydrophobic surfaces \cite{dupuis2005modeling}. We then present our results in section III, where we characterize the three dynamic regimes identified: compact droplet, droplet break up and oscillating droplet. Finally, we summarise the key results and provide an outlook for future work in section IV.

\section{Methods}

\subsection{The Continuum Model}
The fluid equations of motion are described by the combination of the continuity and Navier-Stokes equations. The continuity equation describes the conservation of mass,
\begin{equation}
        \partial_t \rho + {\boldsymbol{\nabla}} \cdot (\rho {\boldsymbol{v}}) = 0, \label{eq:continuity-eq}
\end{equation}
where $\rho$ and $\boldsymbol{v}$ are the fluid density and velocity. The Navier-Stokes equation describes the conservation of momentum for the fluid, and it can be written as
\begin{equation}
       \partial_t (\rho {\boldsymbol{v}}) + {\boldsymbol{\nabla}} \cdot (\rho {\boldsymbol{v}} \otimes {\boldsymbol{v}}) = - {\boldsymbol{\nabla}} \cdot {\boldsymbol{P}} + {\boldsymbol F}_{\rm ext} + {\boldsymbol F}_{\rm s} \label{eq:NSE} + {\boldsymbol{\nabla}} \cdot [\eta ({\boldsymbol{\nabla v}} + {\boldsymbol{\nabla v}}^T)].
\end{equation}
Here, $\eta$ is the dynamic viscosity which depends on the fluid density, ${\boldsymbol F}_{\rm ext}$ is the external force (such as gravity) applied to the system, and ${\boldsymbol F}_{\rm s}$ represents forces due to fluid-solid interactions. The expressions of ${\boldsymbol F}_{\rm ext}$ and ${\boldsymbol F}_{\rm s}$  will be discussed in the next sub-section. $\boldsymbol{P}$ is the pressure tensor of the fluid, and it depends on the thermodynamics of the fluid.

In this work, we use the following free energy to capture liquid-gas coexistence and the liquid-gas surface tension \cite{mazloomi2014entropic, wohrwag2018ternary}
\begin{equation}\label{eq:free-energy-specific}
    E = \int \left[\dfrac{\lambda}{2} \psi_{eos} + \dfrac{\kappa}{2}\,\nabla^2 \rho \right]\ dV,
\end{equation}
where the integration is done over the volume of the simulation domain. $\lambda$ and $\kappa$ are introduced to tune the surface tension (here, we use $\sigma_{lg} = 0.184$ in lattice unit) and the interface width between the liquid and gas phases. The form of the free energy function $\psi_{eos}$ depends on the fluid equation-of-state (EOS). It is related to the bulk pressure of the system by:
\begin{equation}
    p_b= \rho \bigg[ \dfrac{d \psi_{eos}}{d\rho}  \bigg] - \psi_{eos}.
\end{equation}
Here, we will use the Carnahan-Sterling EOS such that
\begin{eqnarray}
    p_b = \rho R T\, \dfrac{1 + (b\rho/4) + (b\rho/4)^2 - (b\rho/4)^3}{(1 - b\rho/4)^3} - a\rho^2, \label{eq:CS-EOS-1} \\
    \psi_{eos} = \rho \Bigg[ C - a \rho - \dfrac{8RT (-6 + b\rho)}{(-4 + b\rho)^2} + RT\,\log\rho \Bigg]. \label{eq:CS-psi}
\end{eqnarray}
Based on the work of W\"{o}hwarg et al. \cite{wohrwag2018ternary}, we choose in lattice units: $a = 0.037$, $b = 0.2$, $R = 1$, and $T_c = 0.3773 a/(bR)$. $C$ is chosen such that $\psi_{eos}(\rho_g) = \psi_{eos}(\rho_l)$, and the coexisting gas ($\rho_g$) and liquid ($\rho_l$) densities can be found by using the standard common tangent construction method. In this model, the temperature $T$ governs the liquid-gas density ratio obtained. Typically, we use $T = 0.61$ in lattice unit, leading to a liquid-gas density ratio of $\rho_l/\rho_g \simeq 800$.

This free energy model will enter the equations of motion via the pressure tensor
\begin{equation}
    {\boldsymbol{\nabla}} \cdot {\boldsymbol{P}} = \rho {\boldsymbol{\nabla}} \mu_{\rho},
\end{equation}
where $\mu_{\rho}= \delta E/\delta\rho$ is the chemical potential. This leads to the following pressure tensor
\begin{equation}\label{eq:binary-pressure-tensor}
    P_{\alpha\beta} = p_b \delta_{\alpha\beta} + \kappa \bigg[ (\partial_\alpha \rho) (\partial_\beta \rho) - \Big[ \rho (\partial_{\gamma\gamma}^2 \rho) + \dfrac{1}{2} (\partial_\gamma \rho)^2 \Big] \delta_{\alpha\beta}  \bigg].
\end{equation}
The bulk pressure, $p_b$, is given by Eq. \eqref{eq:CS-EOS-1} for Carnahan-Starling EOS.

\subsection{The Lattice Boltzmann Method}
We use the entropic lattice Boltzmann method (ELBM) to solve the continuity and Navier-Stokes equations. In ELBM, we introduce a discretised fluid distribution function $f_i (\boldsymbol{x}, t)$ that represents the partial density of fluids with lattice velocity $c_i$ at position $\boldsymbol{x}$ and time $t$. Here we use the so-called D3Q19 lattice, where the weight $(w_i)$ and lattice velocity (${\boldsymbol{c}}_i$) for each direction $i$ are provided in Table \ref{tab:velocity_set}. We then evolve the discretised distribution function using \begin{equation}\label{eq:elbm}
    f_i( {\boldsymbol{x}} + {\boldsymbol{c}}_i \Delta t,\,t + \Delta t) - f_i({\boldsymbol{x}},t) =
    \alpha\beta [ f_i^{eq}(\rho, {\boldsymbol{u}}) - f_i({\boldsymbol{x}},t) ] + \, F_i({\boldsymbol{x}},t).
\end{equation}

The first term on the right-hand side is the collision term. $f_i^{eq}$ is the equilibrium distribution function,
\begin{equation}\label{eq:equilibdistfunction}
    f_i^{eq} (\rho, {\boldsymbol{u}}) =
    w_i \rho \Bigg( 1 + \dfrac{u_\alpha c_{i\alpha}}{c_s^2} + \dfrac{u_\alpha u_\beta \big(c_{i\alpha} c_{i\beta} - c_s^2 \delta_{\alpha\beta} \big)}{2c_s^4} \Bigg),
\end{equation}
where $c_s = 1/\sqrt{3}$ is the standard speed of sound in LBM and $\delta_{\alpha\beta}$ is the Kronecker delta function.  In the collision term, the variable $\alpha$ is the non-trivial root for the discrete entropy function
\begin{equation}\label{eq:elbm-alpha}
    H(f_i' + \alpha[f^{eq}(\rho, {\boldsymbol{u}} + \Delta {\boldsymbol{u}}) - f_i'] ) = H(f_i'),
\end{equation}
with $f'_i = f_i + [f_i^{eq}(\rho, {\boldsymbol{u}} + \Delta {\boldsymbol{u}}) - f_i^{eq}(\rho, {\boldsymbol{u}}) ]$ and $H = \sum_i f_i \ln (f_i/w_i)$. The variable $\alpha$ is typically 2.0 in the bulk liquid and gas phases, equivalent to what it would be if the Bhatnagar-Gross-Krook (BGK) approximation is used \cite{kruger2016lattice}, but its value will vary in a non-trivial manner across the liquid-gas interface. The variable $\beta$ is related to the kinematic viscosity $(\nu=\eta/\rho)$ through
\begin{equation}\label{eq:elbm-nu}
    \nu = \dfrac{(\beta^{-1} -1)\,c_s^2}{2}.
\end{equation}

The second term on the right hand side of Eq. \ref{eq:elbm} is called the forcing term. We use the exact difference method (EDM) whose form is \cite{kupershtokh2009eos}
\begin{equation} \label{eq:EDM}
    F_i = [f_i^{eq} (\rho, {\boldsymbol{u}} + \Delta {\boldsymbol{u}}) - f_i^{eq} (\rho, {\boldsymbol{u}})].
\end{equation}
Following previous works on ELBM \cite{mazloomi2014entropic, wohrwag2018ternary}, the correction term $\Delta \boldsymbol{u}$ is defined as
\begin{equation}
    \rho \Delta{\boldsymbol{u}} = ({\boldsymbol F}_\text{thermo} + {\boldsymbol F}_\text{ext} + {\boldsymbol F}_\text{s}) \, \Delta t.
\end{equation}
${\boldsymbol F}_\text{thermo}$ is the thermodynamic force, defined as ${\boldsymbol F}_{\rm thermo} = {\boldsymbol{\nabla}} \cdot (\rho c_s^2 {\boldsymbol{I}} - {\boldsymbol{P}})$. ${\boldsymbol F}_\text{ext}$ is the external force, where due to gravity, ${\boldsymbol F}_\text{ext} = \rho {\boldsymbol g}$ with ${\boldsymbol g}$ the gravitational acceleration. For the fluid-solid interface force, ${\boldsymbol F}_\text{s}$, we introduce a body force for fluid nodes whose direct neighbours are solid nodes. The form of the force is \cite{bala2019wetting}
\begin{equation}\label{eq:forcing-eq}
    {\boldsymbol{F}}_s ({\boldsymbol{x}},t) = \kappa^w\rho^{rel} ({\boldsymbol{x}}) \sum_i w_i s({\boldsymbol{x}} + {\boldsymbol{c}}_i \Delta t) {\boldsymbol{c}}_i.
\end{equation}
$\kappa^w$ is the fluid-solid interaction intensity, and its value can be varied from negative (non-wetting, where the contact angle is $>90^\circ$) to positive (wetting, where the contact angle is $<90^\circ$). $\rho^{rel} ({\boldsymbol{x}})= (\rho({\boldsymbol{x}}) - \rho_g)/(\rho_l - \rho_g)$ is the rescaled density. The function $s({\boldsymbol{x}} + {\boldsymbol{c}}_i \Delta t)$ takes the value of 1 when the corresponding fluid node is next to a solid node and 0 everywhere else. Here, we use the staircase boundary approximation for the circular cross-section of the solid fibre.

From Eq. (\ref{eq:elbm}), we can reconstruct the fluid density $(\rho)$ and its bare velocity $(\boldsymbol{u})$ as:
\begin{align}
    \rho({\boldsymbol{x}}, t) &= \sum f_i({\boldsymbol{x}}, t), \\
    \rho {\boldsymbol{u}}({\boldsymbol{x}}, t) &= \sum f_i({\boldsymbol{x}}, t)\, {\boldsymbol{c}}_i .
\end{align}
$\boldsymbol{u}$ is related to the actual velocity by $\boldsymbol{v} = \boldsymbol{u} + \Delta\boldsymbol{u}/2$.

\begin{table}[th]
    \centering
    \caption{The weights $w_i$ and lattice velocities $\boldsymbol{c_i}$ for D3Q19 lattice in LBM.}
    \label{tab:velocity_set}
    \begin{tabular}{r|c|c|c|c|c|c|c|c|c|c|c|c|c|c|c|c|c|c|c}
        $i$    & 0 & 1  & 2  & 3  & 4  & 5  & 6  & 7  & 8  & 9  & 10 & 11 & 12 & 13 & 14 & 15 & 16 & 17 & 18 \\ \midrule
        $w_i$ &
        $\frac{1}{3}$ &
        $\frac{1}{18}$ &
        $\frac{1}{18}$ &
        $\frac{1}{18}$ &
        $\frac{1}{18}$ &
        $\frac{1}{18}$ &
        $\frac{1}{18}$ &
        $\frac{1}{36}$ &
        $\frac{1}{36}$ &
        $\frac{1}{36}$ &
        $\frac{1}{36}$ &
        $\frac{1}{36}$ &
        $\frac{1}{36}$ &
        $\frac{1}{36}$ &
        $\frac{1}{36}$ &
        $\frac{1}{36}$ &
        $\frac{1}{36}$ &
        $\frac{1}{36}$ &
        $\frac{1}{36}$ \\ \midrule
        $c_{ix}$ & 0 & +1 & -1 & 0  & 0  & 0  & 0  & +1 & -1 & +1 & -1 & 0  & 0  & +1 & -1 & +1 & -1 & 0  & 0  \\ \midrule
        $c_{iy}$ & 0 & 0  & 0  & +1 & -1 & 0  & 0  & +1 & -1 & 0  & 0  & +1 & -1 & -1 & +1 & 0  & 0  & +1 & -1 \\ \midrule
        $c_{iz}$ & 0 & 0  & 0  & 0  &    & +1 & -1 & 0  & 0  & +1 & -1 & +1 & -1 & 0  & 0  & -1 & +1 & -1 & +1
    \end{tabular}
\end{table}

\subsection{Simulation Setup}
\begin{figure*}[ht]
    \centering
    \includegraphics[width=1.0\linewidth]{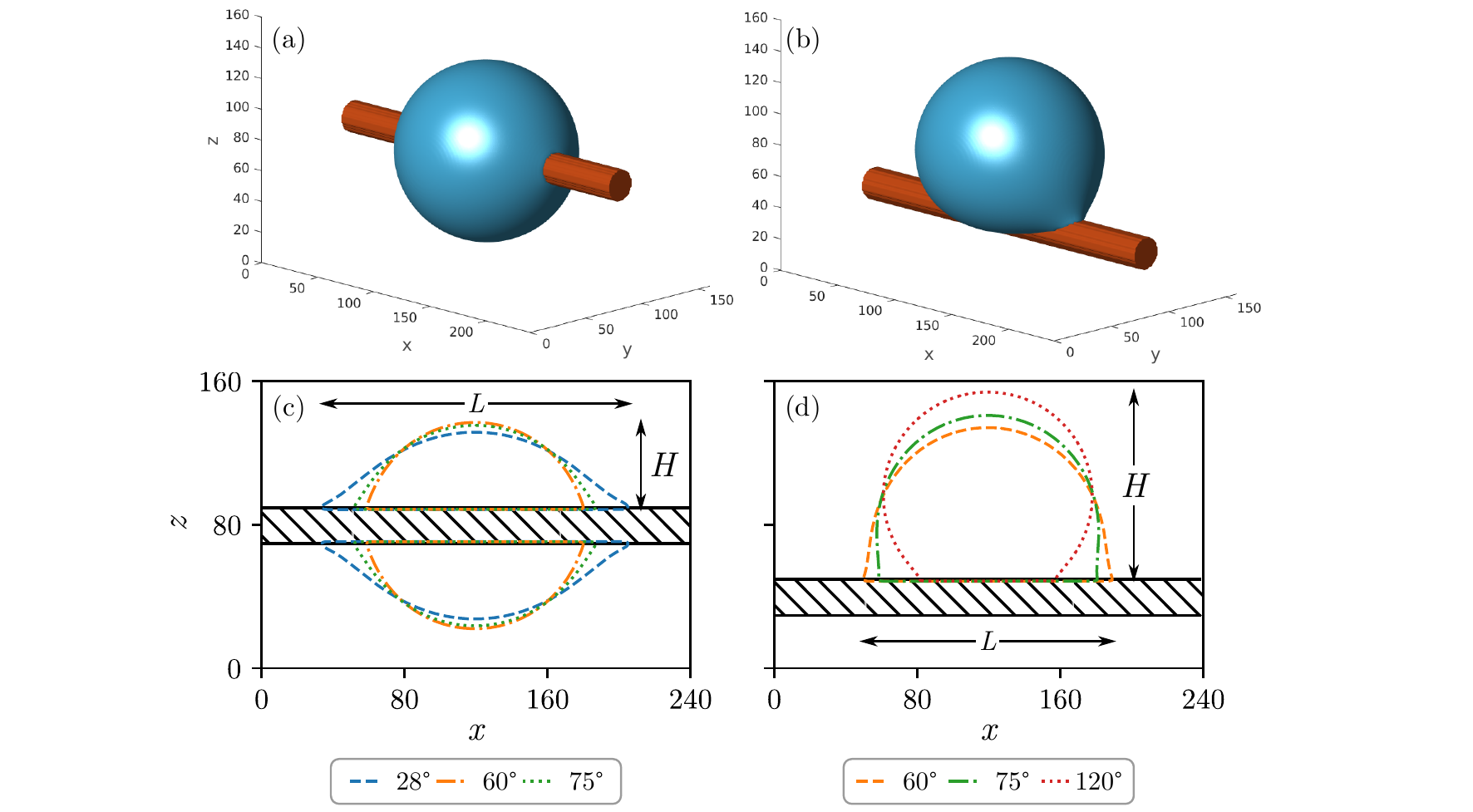}
    \caption{(a) and (b) show the three-dimensional view of a droplet in the barrel and clamshell morphologies, respectively. In both cases, the normalised droplet volume is  $L_\text{R} = 9.28$ and the equilibrium contact angle is $\theta_e = 75^o$. (c) and (d) show the cross-sections of droplets with the same volume $L_R$ but different contact angles $(\theta_e)$. $L$ denotes the droplet length along the fibre, while $H$ denotes the height of the droplet perpendicular to the fibre.
    }
    \label{fig:solid-config}
\end{figure*}

At equilibrium, it is well-known that a droplet on fibre can adopt both barrel and clamshell morphologies, depending on the contact angle $(\theta_e)$ and its volume $(\Omega)$ relative to fibre radius $(r_f)$ \cite{eral2011drops,mchale2002global, chou2011equilibrium, de2012buoyant}. Typically, the barrel and clamshell morphologies are favoured for small contact angles with large volumes and large contact angles with small volumes, respectively. A bistable region is also observed where both morphologies are mechanically stable. Using ELBM, we are able to simulate both barrel and clamshell morphologies, as depicted in Fig. \ref{fig:solid-config}. Fig. \ref{fig:solid-config} (a) and (c) show the droplet in the barrel morphology, while Fig. \ref{fig:solid-config} (b) and (d) show the droplet in the clamshell morphology.   The different lines for the cross-sections in Fig. \ref{fig:solid-config} (c) and (d) correspond to several values of contact angle formed by the droplet on the fibre surface. To characterise the droplet shape, we denote with $L$ the length of the droplet along the fibre axis, and with $H$ the height of the droplet perpendicular to the fibre.

In this work, we are mainly interested in the droplet dynamics as a body force is introduced parallel to the fibre. We identify three non-dimensional control parameters. The first one is the contact angle $\theta_e$ formed by the droplet with the fibre surface. The second one is the relative droplet size (to the fibre radius), defined as
\begin{equation}
    L_R = \dfrac{\Omega^{1/3}}{r_f},
\end{equation}
where $\Omega$ is the droplet volume and $r_f$ is the fibre radius. The third control parameter is the nondimensional body force, represented by the Bond number (Bo),
\begin{equation}
    \text{Bo} = \dfrac{\rho_l g_x \Omega^{2/3}}{\sigma_{lg}},
\end{equation}
where $\rho_l$ is the liquid density, $g_x$ is the gravitational acceleration parallel to the fibre, and $\sigma_{lg}$ is the liquid-gas surface tension.
In addition to the three nondimensional control parameters, the main observable considered in this work is the nondimensional droplet velocity, represented by the Capillary number (Ca),
\begin{equation}
    \text{Ca} = \dfrac{\eta_l v_x}{\sigma_{lg}},
\end{equation}
where $\eta_l$ is the dynamic viscosity of the droplet and $v_x$ is the droplet velocity along the fibre.

There are two main sets of simulation data that we generate. First, we focus on the clamshell morphology and study its stability against satellite droplet formation as we vary the Bond number and relative droplet size. To vary Bo we keep the droplet volume fixed to $\Omega = 8\times10^6$ lattice units (l.u.), and tune the value of $g_x$, obtaining Bo values ranging between 0.842 and 2.221. For simplicity, we set $g_y$ and $g_z$ to 0. To vary the relative droplet size, we keep the volume constant and vary the fibre radius employing the following values: $r_f$ = 10, 12, 16, 20, 24, 30, 40, and 60 l.u.. This leads to relative droplet size values $L_R$ of 9.28, 7.74, 5.80, 4.64, 3.87, 3.09, 2.32, and 1.55, respectively. For the droplet contact angle, we choose the value of $\kappa^w$ in Eq. \eqref{eq:forcing-eq} that leads to $\theta_e = 75^o$ on a flat surface.

In the second set of data, we investigate the scaling law between the droplet velocity and its driving force, comparing barrel and clamshell morphologies.
In this case, we vary the relative droplet size by fixing the fibre radius to $r_f = 10$ l.u., and vary the droplet volume, leading to $L_R = 9.28, 8.43, 6.69, 5.31$. For these values, both barrel and clamshell morphologies are mechanically stable when the contact angle $\theta_e$ is $60^o$ and $75^o$ \cite{eral2011drops}. We also simulate barrel morphologies with $\theta_e = 28^o$ and clamshell morphologies with $\theta_e = 120^o$. For these combinations of volume and contact angles, only one  morphology, respectively barrel and clamshell, is stable. Furthermore, the gravitational acceleration $g_x$ is varied such that we obtain values of Bo ranging between $3.969\times10^{-3}$ and $3.793$.

As is common in lattice Boltzmann simulations, we use the so-called staircase approximation \cite{khirevich2015coarse,liu2015symmetry} to model the fibre circular cross-section. In this approach, a curved or inclined surface is approximated by piecewise facets that follow the structure of the underlying cubic lattice. This staircase approximation introduces roughness to the fibre cross section and this will modify the effective contact angle. Such roughness due to the staircase approximation is not constant along the fibre cross section. However, as a simple estimate, we can apply the Wenzel contact angle equation \cite{wenzel1936resistance}, and given a typical roughness of 1.2-1.3, the deviation is $<10^o$ for all the cases considered here. Furthermore, since we consider droplet motion along the fibre, the effect of contact line pinning due to the staircase approximation is minimal. Such contact line pinning slightly affects the neck dynamics when a satellite droplet separates from the main droplet. While not a focus of this work, it also has an effect on the shape transition between clamshell and barrel morphology when bistability exist since in the transient states contact line motion occurs across the fibre roughness.

In our simulations, we first initialise the droplet on a fibre setting $g_x = 0$ and let it equilibrate until it reaches its static equilibrium shape. Then, a constant gravitational acceleration $g_x >0$ corresponding to the target Bond number is applied instantaneously to the droplet, and the resulting droplet dynamics is monitored. This can lead to steady moving droplets corresponding to deformed barrel and clamshell morphologies, and also non-steady state shapes with periodic shape oscillations or the formation of satellite droplets. Since the instantaneous application of the body force is expected to affect the transient dynamics, we also selectively simulated an incremental increase of the body force, obtaining qualitatively similar results, though quantitative differences are observed for the regime with satellite droplet formation.

\section{Results}
\subsection{Dynamic Regimes for Clamshell Morphology}
\begin{figure*}[th!]
    \centering
    \includegraphics[width=1.0\linewidth]{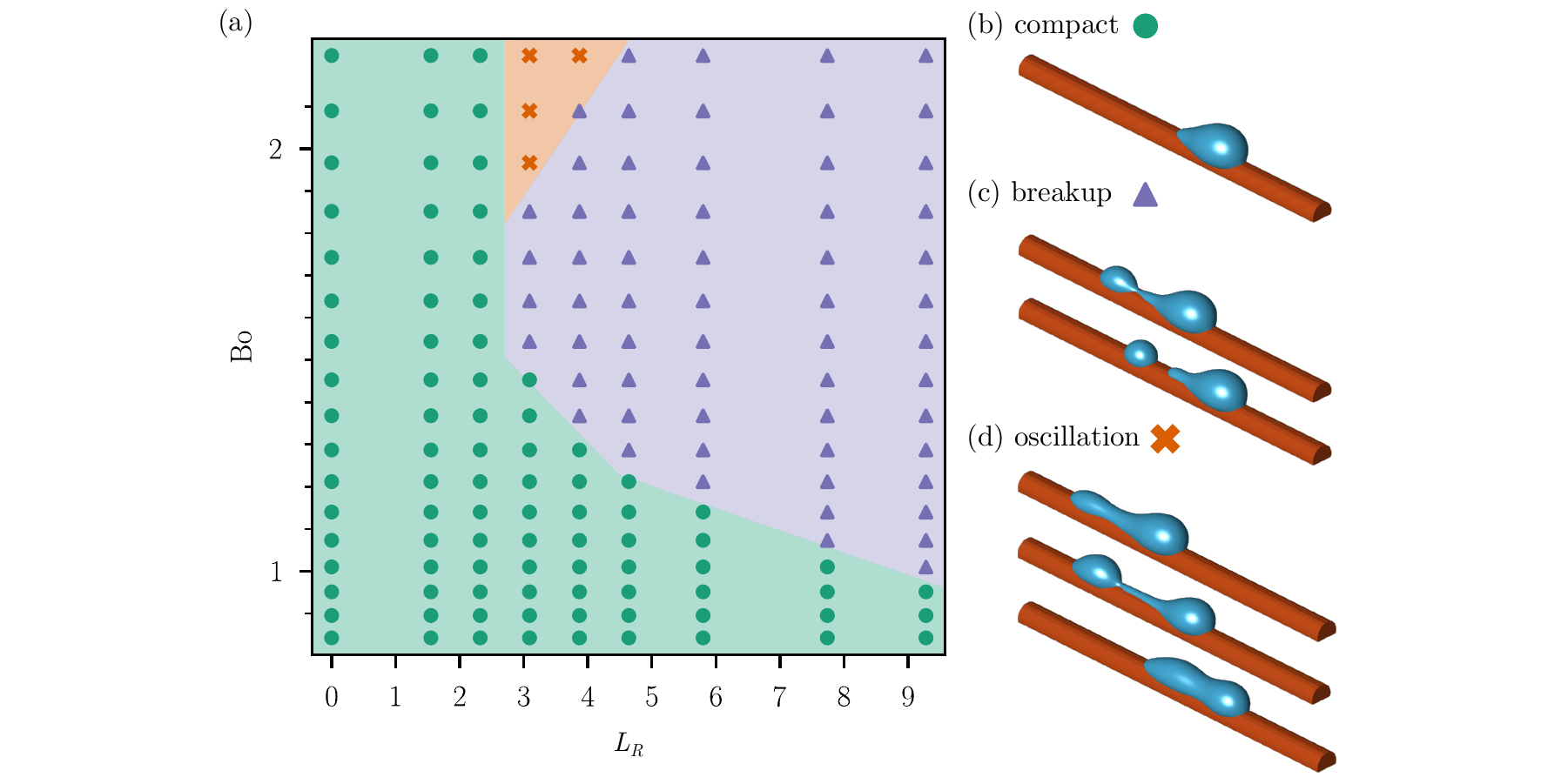}    \caption{(a) The dynamic phase diagram for a droplet in a clamshell configuration for varying Bond number (Bo) and size ratio between the droplet and the fibre radius $(L_R)$. There are three different possible outcomes, and they are illustrated in the subsequent panels for (b) compact droplet, (c) droplet breakup, and (d) oscillating droplet.}
    \label{fig:high-bo-outcome}
\end{figure*}

Fig. \ref{fig:high-bo-outcome}(a) illustrates three dynamic regimes detected in our simulations of a droplet in the clamshell configuration. They form a phase diagram in the parameter space of the Bond number (Bo) and the size ratio between the droplet and the fibre radius, $L_R$. We denominate the three regimes referring to the droplet shape as (i) compact droplet (Fig. \ref{fig:high-bo-outcome}(b)), (ii) droplet breakup (Fig. \ref{fig:high-bo-outcome}(c)), and (iii) oscillating droplet (Fig. \ref{fig:high-bo-outcome}(d)).

The compact droplet regime is highlighted with the circle symbols and a green background. Here, the droplet first undergoes a transient state where it is accelerated, and some shape oscillations are triggered. However, after the oscillations are damped, the droplet will eventually move along the fibre with a steady state shape. In practice, for the classification done in this work, we identify a droplet in the compact regime when: (a) there are no satellite droplets produced by the main droplet; and (b) the residual oscillation triggered by the transient state is damped with a maximum variation of 1\% within the simulation time window. As expected, this regime is found at low Bo number when the droplet is only weakly deformed when compared to its equilibrium shape. In this case, the droplet receding contact line has a rounded shape without any tail. With increasing Bo, and correspondingly the droplet velocity, a visible tail can be observed in the steady state, as illustrated in Fig. \ref{fig:high-bo-outcome}(b). This is reminiscent to previous works on pearling stability on a flat surface \cite{podgorski2001corners,peters2009coexistence,engelnkemper2016morphological}.

The droplet breakup regime is highlighted by the triangle symbols and a blue background. Here, during the transient state, the droplet breaks up leaving behind one or several satellite droplets. The satellite droplets are considerably larger than that observed for the pearling instability studied on flat surfaces \cite{podgorski2001corners,peters2009coexistence,engelnkemper2016morphological}. However, due to the demanding computational resources, we are limited to relatively small droplet size and simulation domain. It will be interesting to study larger systems in the future.
A key observation in our work is that the stability of the compact droplet regime depends on the relative size of droplet and fibre radius. As observed in Fig. \ref{fig:high-bo-outcome}(a), at parity of droplet volume, when the fibre has a large radius of curvature $r_f$ (alternatively, low curvature), the droplet can reach a steady state shape without forming satellite droplets for the entire range of the Bo number explored in this phase diagram. For smaller fibre radius, in contrast, the droplet becomes unstable, leading to the formation of satellite droplets, as shown in Fig. \ref{fig:high-bo-outcome}(c). Interestingly, the onset of this instability shifts to lower Bo as the fibre radius $r_f$ becomes smaller. It is worth noting that, in this work, we assume an idealised fibre where there is no contact angle hysteresis along the droplet. A previous simulation study by Yang et al. \cite{yang2020effects} on a flat surface also highlighted satellite droplet formation, but correlated it with increasing contact angle hysteresis.

In the oscillating droplet regime, the droplet oscillates regularly as it moves along the fibre. Over the range of parameters explored in Fig. \ref{fig:high-bo-outcome}(a), this is observed for the intermediate range of relative droplet sizes. Initially, the droplet dynamics is akin to that for droplet breakup, where the trailing edge bulges and a neck is formed connecting the rear and the front of the droplet, see e.g., Fig. \ref{fig:high-bo-outcome}(d). However, unlike the droplet breakup regime, the neck does not rupture and the rear portion of the droplet merges with its front. Furthermore, we observe the droplet oscillation is not damped and can repeat over multiple cycles in our simulations. Indeed, it appears to continue indefinitely, suggesting that the energy provided by the body force at a constant rate feeds the droplet oscillation as well as the motion of the centre of mass.

\subsection{Dynamics of Compact Droplets}
\begin{figure*}[ht]
    \centering
    \includegraphics[width=1.0\linewidth]{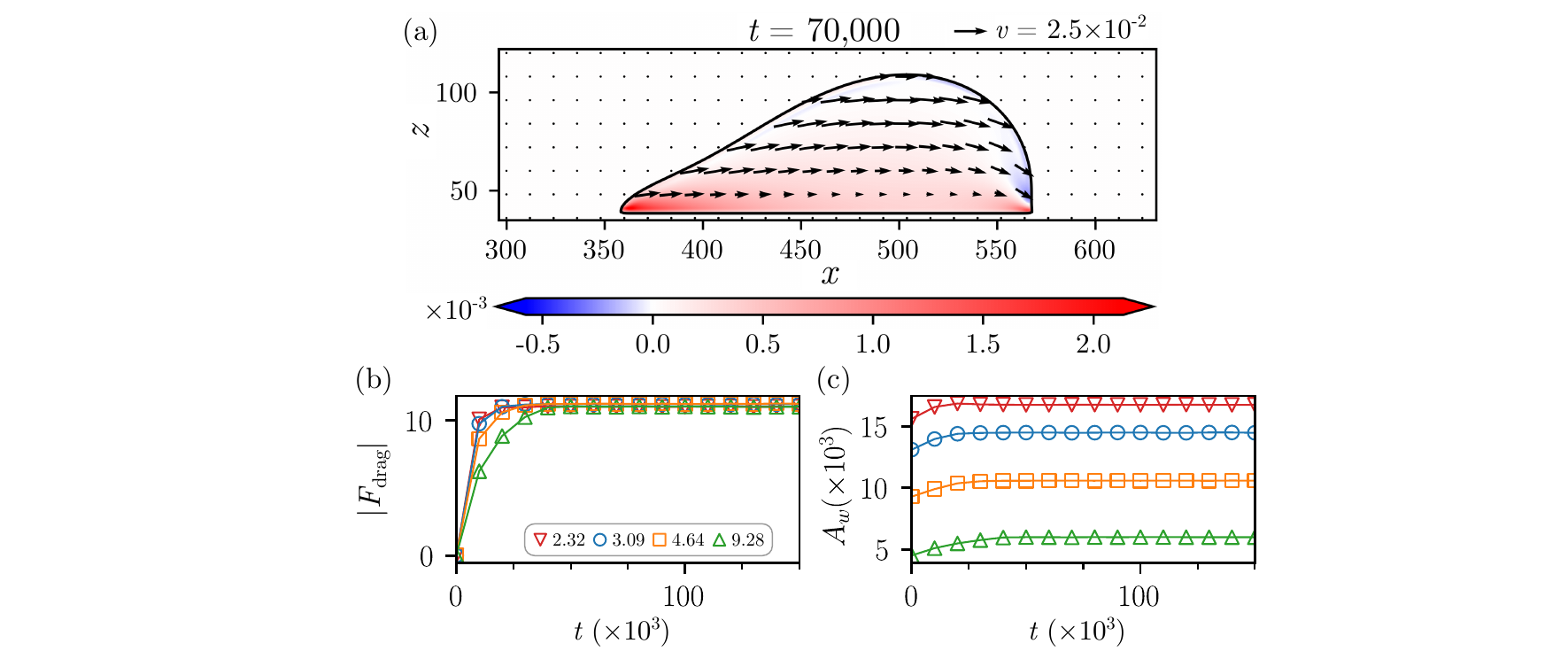}    \caption{(a) Typical droplet flow profile in a compact configuration. The colormap corresponds to the $x$-$z$ component of the strain-rate tensor. Here, we use $L_R = 3.09$ and Bo = 1.453. (b) The time evolution of the solid-liquid friction in the $x$-direction, $F_{\rm drag}$. Different colours denote different values of $L_R=\Omega^{1/3}/r_f$. (c) shows the contact area between the droplet and the fibre $(A_w)$ as a function of time for different normalised fibre curvature. For panels (b) and (c), Bo = 0.842.}
    \label{fig:fig-compact}
\end{figure*}

In this and the next sub-sections, we will discuss the droplet dynamics for each dynamic regime in turn, starting with the compact droplet case. In Fig. \ref{fig:fig-compact}(a) we inspect the middle cross-section of the droplet along the fibre and illustrate its typical flow profile (see the arrows). At steady-state, the internal velocity is dominated by the positive $x$-component with clear variations in the $z$-direction. To characterise this, in Fig. \ref{fig:fig-compact}(a), we have also shown the $x-z$ component of the strain-rate tensor in the plane of the droplet cross-section, defined as $E_{xz} = E_{zx} = (\partial_x v_z + \partial_z v_x)/2$. In our case here, the second term is much larger compared to the first term.

From the strain rate color-map, we can infer that viscous dissipation is dominated by shear close to the solid substrate. Correspondingly, the primary friction force arising at the liquid-solid interface opposes the external body force due to gravity. The liquid-solid friction force in the $x$-direction,  $F_{\rm drag,x}$, can be estimated as:
\begin{equation}
    F_{\rm drag,x} = \int_A \eta_l\,\nabla_\perp v_x \,dA,
\end{equation}
where the integral is over the solid-liquid contact area, $\eta_l$ is the dynamic viscosity of the liquid, and $\nabla_\perp v_x$ is the velocity gradient normal to solid surface.
Fig. \ref{fig:fig-compact}(b) shows the time evolution of the drag force, $F_{\rm drag}$, for several different fibre curvatures. Here, we have chosen Bo = 0.842 where the droplet is in the compact regime for all values of $L_R$ considered in the phase diagram in Fig. \ref{fig:high-bo-outcome}(a). We find that, at steady state, the friction force approaches the same value for all $L_R$ cases. This is expected because we have used the same droplet volume, and at steady state, the friction force must balance the driving force due to gravity. There are, however, two clear variations as we study fibres of different radii. First, the relaxation time to steady state: fibres with smaller curvatures (i.e., larger fibre radii) reach steady state faster than fibres with larger curvatures (smaller fibre radii). Second, the liquid-solid contact area: as shown in Fig. \ref{fig:fig-compact}(c), the contact area is larger for a fibre with smaller curvature (larger fibre radius). Moreover, since the total friction force is the same, as a corollary, the shear stress must be smaller with decreasing curvature (increasing fibre radius).

Next, we investigate in the relation between the droplet velocity and the driving force. Lorenceau and Quer\'{e} \cite{lorenceau2004drops} have previously investigated the spontaneous motion of a perfectly wetting droplet in a barrel configuration on a conical fibre. They argued that, at steady state, the viscous force in the droplet scales as
\begin{equation}
 F_{\rm drag} \sim \eta_l \dfrac{v  \, r_f}{\theta_w},
\end{equation}
where $\eta_l$ is the droplet viscosity, $v$ is the droplet velocity, $r_f$ is the fibre radius, and $\theta_w$ is the effective wedge angle at the contact line. If the wedge angle $\theta_w$ is small ($< 30^o$ or when $H/L < 0.5$) and the droplet moves slowly such that it does not drastically change the shape of the droplet, they further estimated that $\theta_w$ is in the order of the droplet aspect ratio $H/L$, where $H$ is the droplet height perpendicular to the fibre and $L$ is the droplet length along with the fibre. Balancing the viscous friction with the driving force $F_{\rm drive}$, then we obtain
\begin{equation}
    \label{eq:quere-1}
    v \sim \dfrac{H}{L}\,\dfrac{F_{\rm drive}}{\eta_l\,r_f }.
\end{equation}
A similar scaling argument has also been derived by Gilet et al.  \cite{gilet2010droplets}. If the driving force is due to gravity, as it is the case here, the scaling law above can be rewritten in terms of two dimensionless numbers, the Capillary and Bond numbers. Substituting $F_{\rm drive} = \rho_l g_x \Omega$, we can write
\begin{eqnarray}
    \dfrac{\eta_l v}{\sigma_{lg}} &\sim & \dfrac{H}{L} \,\dfrac{\rho_l g_x \Omega^{2/3}}{\sigma_{lg}}\,\dfrac{\Omega^{1/3}}{r_f}, \nonumber \\
    \text{Ca} &\sim & \dfrac{H}{L}\,\dfrac{\Omega^{1/3}}{r_f} \,\text{Bo}, \nonumber \\
    \label{eq:fitting-1}
    \text{Ca} &\sim & \dfrac{H}{L}\,L_R \,\text{Bo}.
\end{eqnarray}
To check the validity of the derived scaling law for partially wetting droplets, we will therefore fit our simulation data against the corresponding equations:
\begin{eqnarray}
    \label{eq:fitting-3}\text{Ca} &=& \alpha_1\left( L_R  \,\text{Bo}\right)^{m_1}, \\
    \label{eq:fitting-4}\text{Ca} &=& \alpha_2\left(\dfrac{H}{L}\,L_R\,\text{Bo}\right)^{m_2},
\end{eqnarray}
where $\alpha_1$, $\alpha_2$, $m_1$, and $m_2$ are numerical constants. The two equations above differ in whether the droplet aspect ratio, $H/L$, is taken into account in the scaling law, as Lorenceau and Quer\'{e} argued it should only be valid when $\theta_w \sim H/L$ is small.

\begin{figure*}[ht]
    \centering
    \includegraphics[width=1.0\linewidth]{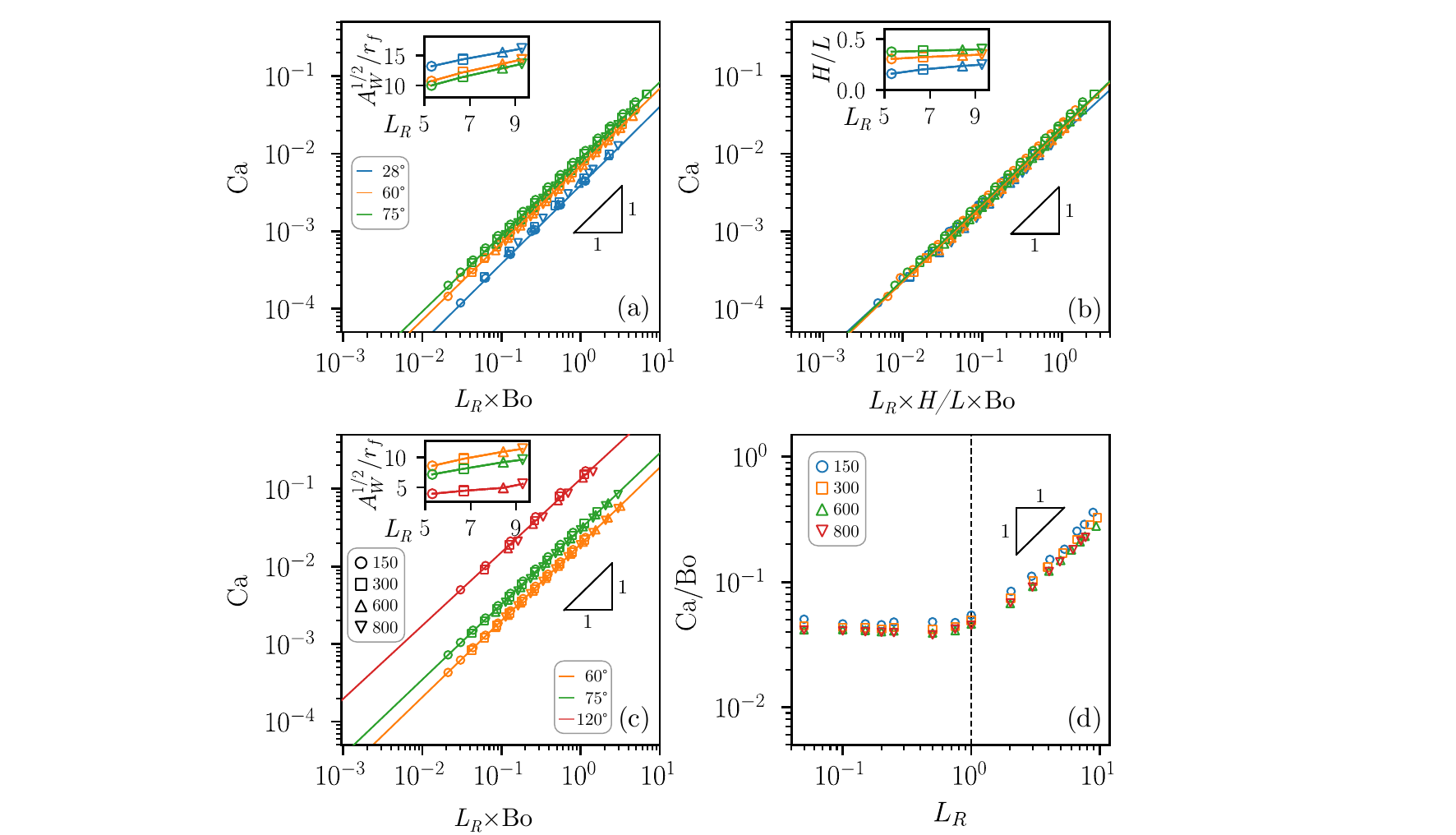}
    \caption{Panels (a) - (c) show the scaling laws between the droplet Capillary (Ca) and Bond (Bo) numbers. The markers are the simulation data points (all panels), while the lines are the best fits based on Eqs. \eqref{eq:fitting-3} (for panels (a) and (c)) and \eqref{eq:fitting-4} (for panel b). Panels (a) and (b) are for the barrel morphology, while panel (c) is for the clamshell morphology. Insets in panels (a) and (c) show the wetted area normalised to the fibre radius $(A_w^{1/2}/r_f)$ as a function of the reduced volume $L_R$. Inset in panel (b) shows the droplet aspect ratio $(H/L)$ as a function of the reduced volume $L_R$. All the data shown in the insets are taken for Bo = 0, and their values do not change significantly with varying Bo across the range considered here. In panel (d), we plot Ca/Bo as a function of reduced volume $L_R = \Omega^{1/3}/r_f$. The dashed line shows $L_R = 1.0$. For each value of the droplet volume $\Omega$, the fibre radius is varied such that $L_R$ ranges from 0.05 to 10.0.}
    \label{fig:bo-ca-graph}
\end{figure*}

Fig. \ref{fig:bo-ca-graph} (a) and (b) show how the simulation data fit against the proposed scaling laws for the barrel morphology. To ensure the scaling law is tested robustly, we have varied the droplet contact angle, the droplet volume, and the applied body force. The fibre radius is fixed at $r_f = 10$. The different colours denote different contact angles $(\theta_e)$, while different symbols correspond to different reduced volumes $(L_R)$. We find that the exponents ($m_1$ and $m_2$) are always very close to 1 (between 0.944 to 1.012) for the range of Bo considered. Deviations from 1 can be observed for larger Bo (typically when Bo > 0.3, though it depends on droplet morphology and contact angle), accompanied by clear deformations of the droplet shape. In this low Bo regime, adding or removing the $H/L$ prefactor does not make the exponents change significantly. Therefore, for a partially wet fibre, the relationship between the velocity and the external force is linear. However, by incorporating the aspect ratio prefactor, all our results collapse into a single line, including fibres with different wettabilities. This observation is well aligned to the argument by Lorenceau and Quer\'{e} that the use of $\theta_w \sim H/L$ in the scaling law is reasonable only if $H/L$ < 0.5 \cite{lorenceau2004drops}. The inset in Fig. \ref{fig:bo-ca-graph}(b) shows that for barrel morphology, $H/L$ is below 0.5.

The linear relation between Ca and $L_R\times$ Bo can also be observed for the clamshell morphology, as demonstrated in Fig. \ref{fig:bo-ca-graph} (c). However, unlike for the barrel morphology, we are unable to collapse all the data points for different contact angles into a single master curve irrespective of whether we include the aspect ratio $H/L$ prefactor (data not shown) or not. Indeed, the aspect ratio $H/L$ is typically larger, $> 0.5$, for the clamshell morphology compared to barrel. Furthermore, in Fig. \ref{fig:bo-ca-graph} (c), the coefficient $\alpha_1$ increases monotonically with $\theta_e$. With increasing contact angle, the droplet contact area (see the inset) is smaller, and the droplet centre of mass is further away from the fibre surface, and as a result, the liquid-solid friction is smaller and droplet velocity is faster. The same argument applies when comparing the barrel and clamshell morphologies. At parity of contact angle and driving force, the clamshell is faster than the barrel because the wet area is smaller.

The data shown in Fig. \ref{fig:bo-ca-graph} (c) correspond to cases where the droplet size is large compared to the fibre radius. We expect that, in the limit of where the droplet is small compared to the fibre, we should recover the scaling law proposed for a droplet moving on an inclined flat surface \cite{kim2002sliding, le2005shape, puthenveettil2013motion}. By balancing the driving gravitational force and dissipative viscous force, Kim {\emph{}{et al.}} derived a relation for the droplet steady-state velocity \cite{kim2002sliding}:
\begin{equation}\label{eq:scaling-flat-1}
    v \sim \dfrac{\rho\,g_x\,\Omega}{\eta_l\,L_\perp},
\end{equation}
where $L_\perp$ is the projection of the drop contact perimeter in the direction orthogonal to the motion. Eq. \eqref{eq:scaling-flat-1} can further be rewritten into:
\begin{align}
   \dfrac{\eta_l v}{\sigma_{lg}} &\sim \dfrac{\Omega^{1/3}}{L_\perp} \, \dfrac{\rho\,g_x\,\Omega^{2/3}}{\sigma_{lg}}, \nonumber \\
   \label{eq:scaling-flat-2}\text{Ca} &\sim \dfrac{\Omega^{1/3}}{L_\perp} \, \text{Bo}.
\end{align}
The key term to consider is the ratio $(\Omega^{1/3}/L_\perp)$ as we vary the size ratio between the droplet and the fibre. When this ratio is small, corresponding to the flat surface limit, for a given contact angle $\theta_e$, $L_\perp$ scales with $\Omega^{1/3}$. This leads to an expected linear scaling between Ca and Bo, without any dependency on the fibre radius. In contrast, when this ratio is large, we expect $L_\perp$ to scale as the radius of the fibre $r_f$, which explains the linear scaling between Ca and $L_R\times$ Bo observed in Fig. \ref{fig:bo-ca-graph} (c).

To verify the crossover, in Fig. \ref{fig:bo-ca-graph} (d), we systematically vary the fibre radius $r_f$ (corresponding to $L_R$ values from 0.05 to 10.0) for four different droplet volumes (indicated by the different colours and symbols), while fixing the gravitational acceleration and droplet contact angle. At low $L_R$ values, we clearly observe a plateau in the ratio between Ca and Bo, while for large $L_R$ values, we again find an additional linear dependence on $L_R$. The latter is the regime where the fibre curvature significantly affects the droplet dynamics. Furthermore, as expected the crossover occurs around $L_R = 1.0$, indicated by the dashed vertical line, where the droplet size becomes comparable to the fibre.

\subsection{Dynamics of Droplet Breakup}
\begin{figure*}[ht]
    \centering
    \includegraphics[width=1.0\linewidth]{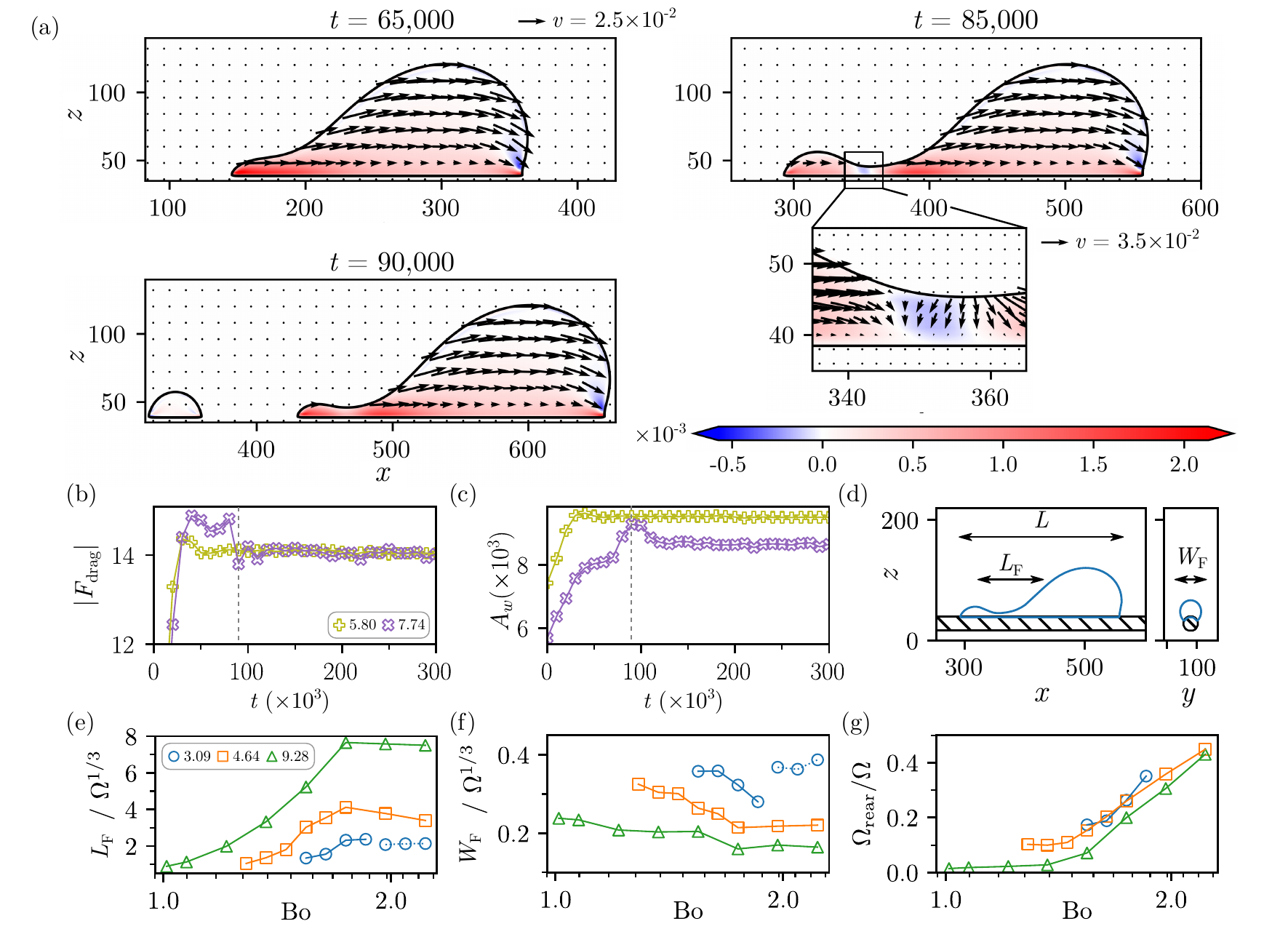}
    \caption{(a)~The droplet flow profile during breakup. The color-map shows the $x$-$z$ component of the strain-rate tensor. Here, Bo~=~1.073 and $L_R = 7.74$. (b)~The drag force $(F_{\rm drag})$ and (c)~the contact area $(A_w)$ between the droplet and the fibre as a function of time for different normalised fibre curvature and Bo~=~1.073. For $L_R = 5.80$, this leads to a compact configuration, while for $L_R = 7.74$, it leads to droplet breakup. Grey dashed line shows the breakup point. (d)~The typical cross-section of the droplet upon breakup, where $L$ denotes the droplet length alongside the fibre, $L$\textsubscript{F} denotes the filament length, and $W$\textsubscript{F} denotes the filament width. (e)~$L$\textsubscript{F} normalised by the droplet characteristic length $(\Omega^{1/3})$, (f)~$W$\textsubscript{F} normalised by $\Omega^{1/3}$, and (g) the rear droplet volume $(\Omega$\textsubscript{rear}) normalised by initial droplet volume $(\Omega)$ as a function of Bond number. In panels (e-g), different colours correspond to different $L_R$, with solid lines for droplet breakup cases and dotted lines for droplet oscillation cases.}
    \label{fig:fig-breakup}
\end{figure*}

During breakup the droplet can be considered as a compound of three main parts: (i) the main droplet body at the front, (ii) a satellite droplet at the rear, and (iii) a liquid filament connecting the main body and the satellite droplet that eventually ruptures. When breakup takes place, the rear side of the droplet starts to bulge (see \emph{t} = 65,000 l.u., Fig. \ref{fig:fig-breakup}(a)), but the fluid flow is predominantly in the $x$-direction, akin to that in the compact droplet regime. As in the previous case, the viscous force is largest close to the solid boundary, as indicated by the color-map illustrating the $x-z$ component of the strain rate. Once the satellite droplet is formed, its volume is typically smaller than the main droplet, so it is subjected by smaller driving force, and its centre of mass moves at smaller velocity than the main droplet. As breakup proceeds, the liquid filament becomes increasingly thinner and more elongated, as illustrated in the snapshot for \emph{t} = 85,000 l.u.. Close to the rupture point in the filament region, we can see a clear downward fluid velocity. This effectively squeezes the neck region leading to droplet breakup, as shown in the snapshot for \emph{t} = 90,000 l.u..

In the previous sub-section, we observed that the value of the drag force converges to the same value for a given Bo, as we vary the fibre radius. Here, when droplet breakup occurs, we find that $F_\text{drag}$ overshoots the value for a steady state compact droplet. Fig. \ref{fig:fig-breakup}(b) and (c) show the drag force $F_{\rm drag}$ and the droplet contact area $A_w$ at Bo~=~1.073 for two cases: $L_R = 7.74$ when the droplet has a compact shape, and $L_R = 5.80$ when the droplet breaks up. Since the droplet contact area is smaller when the curvature is high, we argue the amount of energy which can be dissipated is lower. To compensate this, the droplet will stretch (increasing $A_w$) before the main and satellite droplets separate. After the separation, the total drag force, $F_\text{drag}$, for those two droplets approaches that for a compact droplet.

The droplet breakup mechanism can be characterised by the filament length and the satellite droplet size. Fig. \ref{fig:fig-breakup}(d) shows the cross-section of a typical droplet in the verge of breakup. $L$ denotes the droplet length alongside the fibre, while $L$\textsubscript{F} denotes the filament length. We define the filament boundary as the point where the curvature changes from convex (droplet) to concave (filament).  In Fig. \ref{fig:fig-breakup}(e) and (f), we compare  $L$\textsubscript{F} and $W$\textsubscript{F} at the verge of breakup as a function of Bo for three values of $L_R$: 3.09 ($r_f = 30$), 4.64 ($r_f = 20$), and 9.28 ($r_f = 10$). In general, with decreasing fibre radius, the filament length increases and its width decreases. As the filament elongates over time, it eventually breaks up. Finally, Fig. \ref{fig:fig-breakup}(g) shows the satellite droplet volume ($\Omega$\textsubscript{rear}), measured on the first frame of breakup. We find that the volume of the satellite droplet increases with the Bond number. At the same time, for the same Bond number, higher curvature surface $(1/r_f)$ will induce a smaller satellite droplet volume.

\subsection{Dynamics of Droplet Oscillation}
\begin{figure*}[ht]
    \centering
    \includegraphics[width=1.0\linewidth]{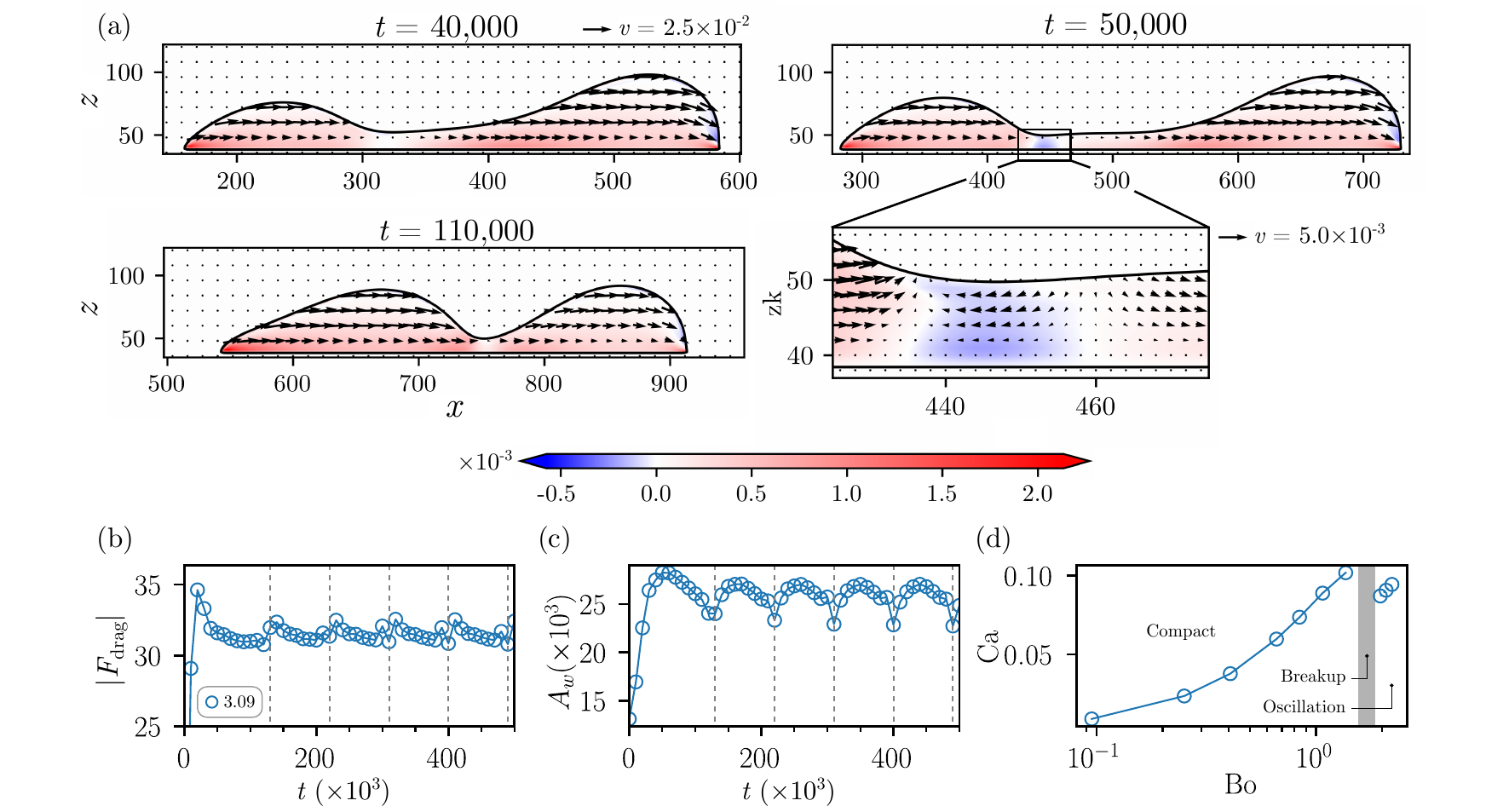}    \caption{(a) The typical droplet flow profile in the droplet oscillation regime. The colormap shows the $x$-$z$ component of the strain-rate tensor. (b) The wetted area and (c) the drag force as a function of time for $L_\text{R} = 3.09$ and Bo = 2.221. (d) Ca as a function of Bo for $L$\textsubscript{R}~=~3.09. For the compact cases, the velocity is measured at steady state condition. For the droplet oscillation cases, the velocity is averaged over once cycle of stretching and contraction.}
    \label{fig:fig-oscillation}
\end{figure*}

In addition to droplet breakup, we also find another dynamic regime at higher Bond number that we term droplet oscillation. This regime is similar to the one described by Yang et al. \cite{yang2020effects} on a flat surface, as illustrated in Fig. \ref{fig:fig-oscillation}(a). Here, in the time range between \emph{t} = 40,000 l.u. and \emph{t} = 50,000 l.u., we observe droplet dynamics similar to the one leading to droplet breakup, where the filament stretches and a downward fluid displacement in the neck region is initiated. However, breakup is not triggered, and the filament survives for a much longer time until eventually the filament shortens (\emph{t} = 110,000 l.u.), the two parts of the droplet recombine, and the cycle is repeated.

In our analysis of the breakup mechanism, we find that smaller fibre curvature (larger fibre radius) will lead to a shorter filament ($L$\textsubscript{F}), wider filament ($W$\textsubscript{F}), and a larger satellite droplet volume ($\Omega$\textsubscript{rear}) (see Fig. \ref{fig:fig-breakup}(e)--(g)) . In Fig. \ref{fig:fig-breakup}(e) and (f), we also include data points for the oscillation cases, which are denoted by dotted lines. Compared to the breakup cases (see the full and dashed blue lines), we find that $L$\textsubscript{F} does not differ significantly, while $W$\textsubscript{F} is clearly larger. In general, filaments with higher aspect ratio ($W$\textsubscript{F}/$L$\textsubscript{F}) are more stable. Furthermore, when the satellite and main droplets are of similar size, they are affected by a similar driving force, and consequently the velocity difference between them is small. Taken together, these limit the extension of the filament region, and the short separation makes it easier for the satellite droplet to recombine with the main droplet in the droplet oscillation cases. This consideration also qualitatively explains why droplet oscillation is detected in a narrow boundary region between compact and breakup in our phase diagram of Fig. \ref{fig:high-bo-outcome}(a). If the fibre curvature is too small, filament formation is suppressed, leading to a stable compact droplet. However, if the fibre curvature is too large, (1) the rear droplet volume will be too small, (2) the filament will be too long, and (3) the contact area will also be too small for the filament to be stable; leading to droplet breakup.

In Fig. \ref{fig:fig-oscillation} (b) and (c), we also show the time evolution of the drag force, $F_{\rm drag}$, and the droplet contact area, $A_w$. A period in these curves corresponds to one cycle of droplet stretching and contraction, and we observe that the droplet oscillation indefinitely repeats over multiple cycles in our simulation. The minimum contact area coincides with the maximum drag force, as shown by the dashed grey lines in Fig. \ref{fig:fig-oscillation}(b) and (c).

Finally, Fig. \ref{fig:fig-oscillation}(d) shows the droplet centre of mass velocity (nondimensionalised into Ca) as a function of Bo when $L$\textsubscript{R}~=~3.09. For the droplet oscillation cases, we plot the average centre of mass velocity over one cycle of stretching and contraction. As we have previously shown in Fig. \ref{fig:bo-ca-graph}, Ca increases proportionally with Bo (for Bo < 1.4) when the droplet has a compact shape. However, as we enter the droplet oscillation regime, we observe that the average Ca is lower even though the external driving force is larger. This reflects the fact that the droplet shape oscillation leads to extra dissipation during droplet motion. With the same droplet oscillation regime, the average Ca increases monotonically again with Bo.

\section{Conclusion}
In this work we have used the lattice Boltzmann method to investigate the dynamics of droplets moving along a fibre, focusing in particular on the clamshell morphology and the partially wetting droplet case. By systematically varying the droplet Bond number and fibre radius, we observed three regimes in the droplet dynamics. First, at low Bond number, the droplet maintains a compact shape. Here, the driving force is balanced by viscous force such that the droplet reaches a steady state. The viscous dissipation is dominated by shear close to the solid substrate, and for smaller fibre radius, the shear stress is larger while the liquid-solid contact area decreases. Analysing the droplet velocity in the low Bond number regime, we further identified a scaling law relating the Capillary number, the Bond number and the size ratio between the droplet and the fibre: $\text{Ca} \propto \text{Bo} \times L_R$ when the fibre curvature is dominant, $L_R > 1.0$; and $\text{Ca} \propto \text{Bo}$ when $L_R < 1.0$ as we approach the flat surface limit. A similar scaling law was further observed for a droplet in the barrel morphology. In fact, for barrel configuration,  all the data can be collapsed into a single line by including a droplet geometrical factor, such that $\text{Ca} \propto  \text{Bo} \, (L_\text{R}) \, (H/L)$.

For large Bond number, the droplet shape becomes unsteady. The dominant unsteady regime is droplet breakup, where satellite droplets are formed at the rear of the moving droplet. Importantly, the transition to the droplet breakup regime strongly depends on the fibre curvature. The larger the curvature (the smaller the fibre radius), the lower the Bond number for the transition between compact droplet and droplet breakup regimes. We rationalised this by characterising the filament prior to satellite formation. The filament is longer and narrower with increasing fibre curvature, which suggests it becomes less stable. Another unsteady regime is droplet oscillation whereby the droplet extends and contracts periodically. This regime was observed in the mid-range of fibre radius explored in this work, indicating subtle dynamics are at play. Compared to the droplet breakup regime, the rear bulge is larger and the filament is wider during the extension phase. As a result, the rear bulge can catch up with the filament and the droplet as a whole contracts. Furthermore, compared to the compact droplet regime, the shape oscillation leads to extra dissipation during droplet motion. Correspondingly, the droplet moves slower in this regime.

Our work highlights several avenues for further investigations. First, we have idealised the fibre geometry in that there is no contact line pinning and contact angle hysteresis along the fibre. In contrast, real fibres unavoidably have some roughness, and previous works on flat surfaces suggest contact angle hysteresis is an important factor in satellite droplet formation \cite{peters2009coexistence, yang2020effects}. Second, we have demonstrated that the substrate curvature can have a strong effect on the resulting droplet dynamics. It will be interesting to explore this concept further on more complex curved geometries, such as a conical fibre or an undulating egg-box substrate \cite{mccarthy2019dynamics, van2021capillary}. Third, we currently limit our work to an external force (e.g. gravity) parallel to the fibre axis. In many practical applications, it will be relevant to consider a driving force that acts on different angles with respect to the fibre axis. It will also be valuable to consider other types of driving forces, such as due to an imposed air flow \cite{sahu2013blowing, davoudi2016barrel, bintein2019self}.

\section*{Acknowledgements}
RC is supported by Durham Global Challenges Centre for Doctoral Training. YR is supported by Kemdikbudristek, RI (Basic Research Grant no. 3490/LL3/KR/2021 (0207/UP-WR3.1/PJN/IV/2021). We thank SJ Avis, F Oktasendra, and JR Panter for useful discussions. This work made use of the facilities of the Hamilton HPC Service of Durham University.

\bibliography{references}

\clearpage
\end{document}